\documentclass{PoS}

\title{AGN and star formation activity in local luminous and ultraluminous infrared galaxies (review)}

\ShortTitle{AGN and star formation activity in local LIRGs and ULIRGs (review)}

\author{\speaker{Almudena Alonso-Herrero}\thanks{Augusto G. Linares Senior
    Research Fellow}\\
        Instituto de F\'{\i}sica de Cantabria, CSIC-Universidad de
        Cantabria, 39005 Santander, Spain\\
        E-mail: \email{aalonso@ifca.unican.es}}

\abstract{The enormous amounts of infrared (IR) radiation emitted by luminous infrared galaxies
  (LIRGs, $L_{\rm
  IR}=10^{11}-10^{12}\,{\rm L}_\odot$) and ultraluminous infrared
galaxies  (ULIRGs,  $L_{\rm
  IR}>10^{12}\,{\rm L}_\odot$) are  produced by dust heated by
intense star formation  (SF) activity and/or an active galactic nucleus
(AGN). The elevated star formation rates and high AGN incidence in
(U)LIRGs make them ideal candidates to study the interplay between SF and AGN
activity in the local universe. In this paper I review recent results
on the physical extent of the SF activity, the AGN detection rate
(including buried AGN), the AGN bolometric contribution to
the luminosity of the systems, as well as the evolution of local
LIRGs and ULIRGs.  The main emphasis of this review is on recent
results from IR observations.
}

\FullConference{Nuclei of Seyfert galaxies and QSOs - Central engine \& conditions of star formation\\
                 November 6-8, 2012\\
                 Max-Planck-Insitut für Radioastronomie (MPIfR), Bonn,
                 Germany}

\begin{document}

\section{Introduction}\label{sec:intro}
Luminous and Ultraluminous infrared (IR) galaxies (LIRGs and ULIRGs)
are defined as having IR
($8-1000\,\mu$m) luminosities in the ranges $L_{\rm
  IR}=10^{11}-10^{12}\,{\rm L}_\odot$ and $L_{\rm
  IR}=10^{12}-10^{13}\,{\rm L}_\odot$, respectively. These immense IR
luminosities are  produced by dust heated by
intense star formation (SF) activity and/or an active galactic nucleus
(AGN)  (see the review of \cite{Sanders1996}). The classical optical study of
\cite{Veilleux1995} and the later   work of \cite{Yuan2010}, among
  others, showed
 that the fraction of sources containing an
optically identified AGN increases with the IR luminosity of the
system. Seyfert nuclei are optically identified in   
$20\%-30\%$ of local LIRGs and up to $\sim 58\%$ of  the most
luminous ULIRGs (see \cite{Yuan2010} and references therein). 

The triggering of the activity of 
local ULIRGs and likely the most luminous local LIRGs is mostly driven by major
mergers and galaxy interactions. Interestingly,
the optical work of \cite{Yuan2010} found that interacting (U)LIRGs
classified as composite ($\equiv$ powered by both SF and AGN) dominate in the
early stages of the interactions, whereas Seyfert nuclei tend to be more common in
the later stages.  This provides a strong support to the evolutionary scenario where a
highly luminous AGN (or even a quasar) appears  in the late
stages of the interaction after the (U)LIRG phase proposed by
\cite{Sanders1988} almost twenty five years ago. However, at the low
luminosity end  ($L_{\rm  IR}<2-3\times 10^{11}\,{\rm L}_\odot$) of 
the local LIRG population major mergers do not dominate in numbers
(\cite{Sanders2004}), and therefore the activity of these LIRGs is
likely triggered by other processes, such as minor mergers, fly-by
companions, and/or secular evolution.

The IR spectral range, and in particular the mid-IR spectral range
($\sim 5-40\,\mu$m), is rich in continuum and spectral features that
can be used to trace both 
AGN and SF activity with the added advantage that in most of this
range the effects of extinction are greatly reduced. 
While optical studies allowed for tremendous progress in
our understanding of the powering mechanisms of (U)LIRGs, they 
may be missing some of the most obscured AGNs likely to
live in these dusty environments. Indeed,  
the  observed deep (in absorption) $9.7\,\mu$m silicate features of 
many local
ULIRGs betray very embedded nuclear sources (see for instance
\cite{Spoon2007}) with  
(minimum) optical extinctions of up to $A_V \sim 60\,$mag (if a
dust screen model is assumed) or even higher for mixed geometries (see
also \cite{Genzel1998}). Local LIRGs, 
while not as extreme,  still show optical extinctions of up to a
few tens of magnitudes (see e.g., \cite{AAH06,Pereira2010LIRGs,AAH12}). 
Thus, we need to identify buried AGN in local (U)LIRGs before we
attempt to  estimate accurately the relative bolometric contributions of SF and AGN
activity to their luminosities. In this paper I 
review briefly recent results using IR observations to quantify the
AGN and SF activity contribution to the energetics of local (U)LIRGs. I  also
discuss  the 
influence of interactions and secular evolution in such activities.

\section{Star Formation Activity in LIRGs and ULIRGs}\label{sec:SFactivity} 
Local (U)LIRGs are known to have very high star formation rates
(SFR). If all their IR was to be produced
only by SF, their IR luminosities would imply SFRs of
$17-170\,{\rm M}_\odot \, 
{\rm yr}^{-1}$ and $170-1700\,{\rm M}_\odot \,
{\rm yr}^{-1}$  for LIRGs and ULIRGs, respectively, using the
\cite{Kennicutt1998} prescription for a Salpeter IMF. 
The most commonly used mid-IR tracers of SF are the [Ne\,{\sc
  ii}]$12.81\,\mu$m fine structure line sometimes combined with the 
[Ne\,{\sc iii}]$15.56\,\mu$m line and 
the polycyclic aromatic features (PAHs). The luminosity of the neon
lines is directly
proportional to the number of ionizing photons (\cite{Roche1991,
  Ho2007}) and thus it probes the most
massive and young stars (ages of less than approximately
10\,Myr). Although the
PAH features probe longer times scales (a few tens of millions of
years), their luminosities are found to correlate well with the IR
luminosity of the system in high metallicity starbursts and ULIRGs
(\cite{Brandl2006, Farrah2007}) indicating that they can be used to
derive SFRs.  Additionally, extended continuum
mid-IR emission, probing warm dust (typically with temperatures of a
few hundred K), can be 
used to estimate the sizes and luminosities of  star forming
regions. 

There is plenty of evidence showing that a large fraction of local
LIRGs have SF extending over scales of a few kpc and not
only in the nuclear regions. For
instance, \cite{Pereira2010LIRGs} used the mapping capability of the
InfraRed Spectrograph (IRS) on-board {\it Spitzer} to observe a small
sample of local LIRGs drawn from the volume-limited sample of
\cite{AAH06}. They demonstrated that these LIRGs show extended
[Ne\,{\sc ii}], [Ne\,{\sc iii}], and PAH 
emission, with the $11.3\,\mu$m PAH emission appearing more extended
than the [Ne\,{\sc ii}] one (see also
\cite{DiazSantos2010a,DiazSantos2011}). This 
was explained because the $11.3\,\mu$m PAH emission can be excited by
stars older 
than those responsible for the neon emission and/or because of the presence of
diffuse PAH emission in photo-dissociation regions.

\cite{DiazSantos2010b} studied the extended mid-IR continuum emission of a large
sample of local (U)LIRGs belonging to the
Great Observatories All-Sky LIRG Survey (GOALS,
\cite{Armus2009}). They found that in a large fraction of local LIRGs
 more than 50\% of their mid-IR emission arises from circumnuclear
 regions extending up to scales of about 10\,kpc (see also \cite{AAH13}). This is in contrast
 with the most luminous LIRGs ($L_{\rm  IR}>6\times 10^{11}\,{\rm
   L}_\odot$)  and ULIRGs where most of their
 mid-IR emission comes from the nuclear regions on scales of less than
 1\,kpc (see also \cite{Soifer2000}). As we shall see in
 Section~\ref{sec:AGNactivity}, this does not 
 necessarily imply that local ULIRGs do not have extended SF
 (see e.g., \cite{RodriguezZaurin2011}), but
 rather that their mid-IR emission
has a large or even a dominant contribution from an AGN.

\section{AGN Activity in LIRGs and ULIRGs}\label{sec:AGNactivity}  
\subsection{Mid-IR AGN indicators}
The mid-IR spectral range is rich in continuum and line features that
can be used to identify and quantify the AGN emission in galaxies. The
most direct way to detect an AGN in this range is by
detection of  fine-structure lines whose
ionization potentials are sufficiently high that they can only be
excited by AGN-produced photons. Among them,  the [Ne\,{\sc v}] lines at
14.32 and $24.32\,\mu$m are the best choice as they are unambiguously
produced by AGN  (see \cite{Genzel1998}). However, a caveat to keep in
mind is that these [Ne\,{\sc v}] lines are not always detected in
local bright AGN (see e.g.,
\cite{Pereira2010lines}). Indeed, the [Ne\,{\sc v}] detection rate
remains relatively low in local LIRGs (approximately 20\%,
\cite{Petric2011,AAH12}) and varies between 25\% (\cite{Veilleux2009})
and $\sim 30-50\%$ (\cite{Farrah2007}) in local ULIRGs 

Other fine-structure lines with intermediate ionization
potentials, such as the [O\,{\sc iv}]$25.89\,\mu$m line, while 
excited by both SF and AGN, they can also be useful when combined with low
ionization lines produced by SF (for instance, the [Ne\,{\sc ii}]
line). The [O\,{\sc iv}] 
line has a higher detection rate than the [Ne\,{\sc v}] lines in
(U)LIRGs, but it is detected in some (U)LIRGs optically classified as
star-forming galaxies. Nevertheless, most local LIRGs
show [O\,{\sc iv}]/[Ne\,{\sc ii}] ratios similar to those of high
metallicity starburst galaxies, with only a small fraction ($\sim
10\%$, mostly bright Seyfert nuclei) showing ratios comparable to those of {\it pure} AGN
(\cite{Petric2011,AAH12}).  ULIRGs on the other hand, show a broader
spread in fractional AGN and starburst contributions to the [O\,{\sc
  iv}] line (\cite{Farrah2007,Veilleux2009}), indicating higher AGN
bolometric contributions as we shall see in Section~\ref{subsec:AGNbol}.

\begin{figure}
\hspace{3cm}
\includegraphics[width=.6\textwidth]{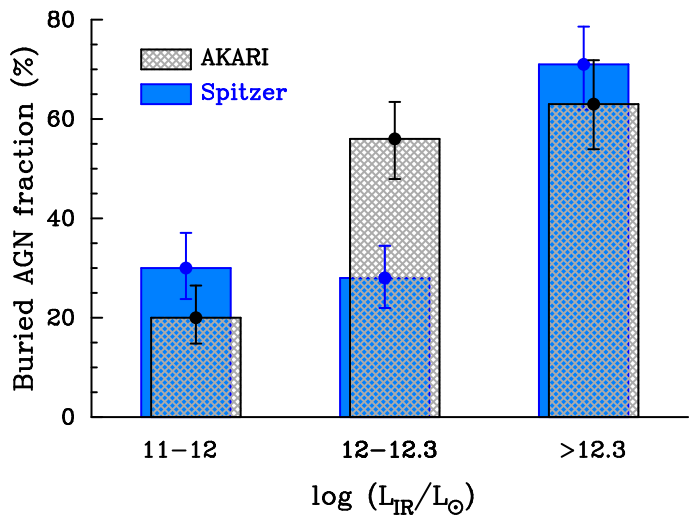}
\caption{Fraction of optically elusive (i.e., not classified 
  as Seyfert nuclei in the optical) buried AGN for local
  LIRGs and ULIRGs. {\it Spitzer} results for local LIRGs are from
  \cite{AAH12}, whereas the {\it Spitzer} results for local ULIRGs and the
  {\it AKARI} results are from \cite{Imanishi2010}. The error bars are the
  1$\sigma$ uncertainties of the fractions based on the number of
  AGN detections relative to the total number of objects 
in each IR luminosity bin.}
\label{fig1}
\end{figure}

As discussed in Section~\ref{sec:SFactivity}, PAH emission is ubiquitous and
remarkably similar in
high metallicity starburst galaxies in the local universe
(\cite{Brandl2006}). Thus, the equivalent widths (EWs) of
different PAH features (e.g., at 3.3, 6.2, 7.7, and $11.3\,\mu$m) are
commonly used to identify AGN especially in composite sources as well
as to detect buried AGN (see e.g., \cite{Imanishi2007, 
Imanishi2010} and references therein). This
is because in the presence of SF, the hot and warm dust
continuum produced by an AGN would effectively decrease the EW of PAH
features. Also it is even possible that the hard radiation field of AGN might destroy the PAH
molecule carriers, although this is still a matter of controversy.  
Therefore, diagnostic
diagrams combining  the
EW of a PAH feature with for instance the [O\,{\sc iv}]/[Ne\,{\sc
  ii}] ratio or the strength of the $9.7\,\mu$m feature are commonly
used to assess the AGN and SF contributions of (U)LIRGs
and galaxies in general 
(see e.g.,
\cite{Genzel1998,Spoon2007,Farrah2007,Veilleux2009,Petric2011,
HernanCaballero2011,AAH12}). 

Hinging on the similarity of the PAH emission in starburst galaxies,
spectral decomposition methods of {\it Spitzer}/IRS spectra have 
demonstrated to be highly effective to
identify AGN and quantify the AGN emission in local (U)LIRGs
(see \cite{Nardini2008,Nardini2010,AAH12}) 
and in high-$z$ IR bright galaxies (e.g.,
\cite{Sajina2007,HernanCaballero2009}). Using the entire {\it
  Spitzer}/IRS spectral 
range ($\sim 5-40\,\mu$m) \cite{AAH12}
detected an AGN component in half  of the volume-limited sample of
local LIRGs of \cite{AAH06}. When also adding those objects optically classified
as composite but without 
a mid-IR AGN detection, \cite{AAH12} found that the AGN detection rate
in local LIRGs is 
$60\%-65\%$ with no dependence with the IR luminosity of the system.
  For a sample of local ULIRGs and using the $5-8\,\mu$m
spectral range \cite{Nardini2010} obtained a similar AGN detection
rate ($70\%$). The AGN detection rate in local (U)LIRGs is much higher
than for instance in the optically selected RSA sample ($5-16\%$, see
\cite{Maiolino1995}). This already hints at a close relation 
between SF and AGN activity in local (U)LIRGs,  
as we shall discuss in Section~\ref{sec:discussion}.

As we have seen, IR observations are very effective in identifying
AGN in the dusty environments of local (U)LIRGs, even in those cases
when the AGN had not been detected optically (also known as optically 
elusive buried AGN). \cite{Imanishi2010} (and references therein) 
recently quantified the
fraction of buried AGN 
in local (U)LIRGs using both {\it Spitzer} mid-IR spectroscopy and
{\it AKARI} near-IR spectroscopy.
Figure~\ref{fig1} summarizes their
results using {\it Spitzer} and {\it AKARI} spectroscopy. We 
also included in this figure  the buried
AGN fraction from {\it Spitzer} 
spectroscopy of local LIRGs from \cite{AAH12}. Except for the 
$\log ({\rm L_{IR}/L}_\odot)= 12-12.3$ bin where there is a 
discrepancy, the {\it AKARI} and {\it Spitzer}
estimates are consistent with each other.
Clearly this fraction
increases with the IR luminosity of the system going from $\sim25\%$
in LIRGs to about $70\%$ in the most luminous ULIRGs. This likely reflects
the fact that the active nuclei of ULIRGs are much more embedded, since
the total AGN fractions of LIRGs and ULIRGs are similar.

\subsection{AGN bolometric contributions}\label{subsec:AGNbol}
Although the combined optical+IR AGN detection rate of local LIRGs is
only slightly less than that of ULIRGs, it is important to quantify
the AGN bolometric contributions in both types of galaxies. One
advantage of the spectral decomposition methods is that they can
provide an estimate of the AGN bolometric contribution to the IR
luminosity of the systems either by using torus models to reproduce
the AGN spectral features in the mid-IR (see \cite{AAH12} for details) or by
making a bolometric correction at a specified mid-IR rest-frame
luminosity of the AGN (as done by \cite{Veilleux2009} and
\cite{Nardini2008,Nardini2010}). \cite{AAH12} found that AGN only 
contribute $5^{+8}_{-3}\%$ of the IR luminosity of local LIRGs. \cite{Petric2011}
found a similar result for the LIRGs in the GOALS sample. These
estimates prove that in
local LIRGs the bulk 
of the IR luminosity is due to SF activity. In local
ULIRGs, the AGN contribution is on average $\sim 30\%-40\%$ depending
on the sample and methods used for the estimates
(\cite{Veilleux2009,Nardini2010}).  We also note here that \cite{Veilleux2009}
found that the AGN bolometric contributions to the luminosity of the system
of some local ULIRGs can be as high as those of Palomar-Green
(PG) quasars. The PG quasars have, however, lower IR luminosities than
ULIRGs.

\begin{figure}
\hspace{2cm}
\includegraphics[width=.4\textwidth,angle=-90]{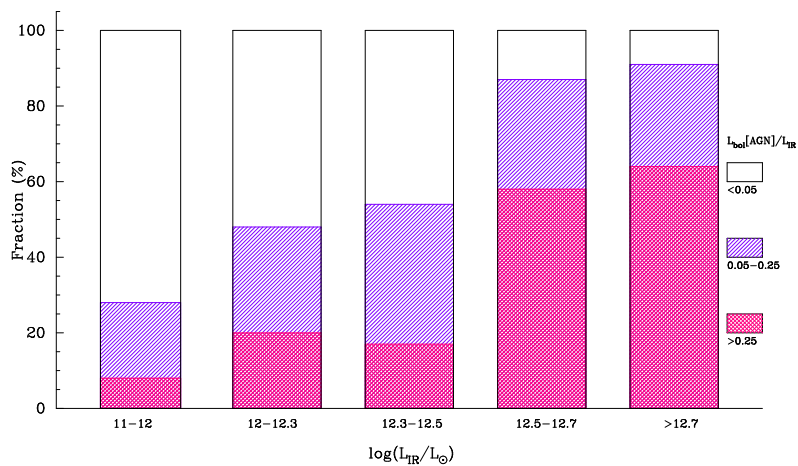}
\caption{Summary of the fractional AGN contribution to the IR
  luminosity of the volume-limited sample of local LIRGs of
  \cite{AAH06,AAH12} and comparison with local ULIRGs of
\cite{Nardini2010}. For each luminosity bin we show three different
ranges of AGN to IR luminosity ratios. Adapted from \cite{AAH12}.}
\label{fig2}
\end{figure}

Figure~\ref{fig2} shows the comparison of AGN bolometric contributions
of local LIRGs from \cite{AAH12} and ULIRGs from \cite{Nardini2010} in
three ranges of bolometric contributions. As can be seen from this
figure, only about $10\%$ of local LIRGs have significant AGN
bolometric contributions (above 25\%). These high AGN bolometric
contributions  increase noticeably with the IR luminosity of ULIRGs (see also
\cite{Veilleux2009}) raising to $\sim 60\%$ for the most luminous ULIRGs.  
\cite{Veilleux2009} also studied the AGN bolometric contributions of
ULIRGs in terms of properties such as IR colors, optical class, and
interaction properties. They found that the AGN contribution is higher
in ULIRGs with ``warmer'' {\it IRAS} colors than those with ``cool''
colors. ULIRGs optically classified as
H\,{\sc ii} tend to show the lowest AGN contributions, while the AGN
fractional contributions increase from those classified as Seyfert 2
to the highest contributions for nuclei optically classified as
Seyfert 1. Finally, there is a slight tendency for increased AGN
contributions at the smallest nuclear separations and latest stages
of the interaction.

\section{Evolution of SF and AGN activity in 
local LIRGs and ULIRGs}\label{sec:discussion} 
Mergers of gas-rich galaxies are efficient in both producing
elevated SFR and transporting gas to the vicinity of the nucleus to allow
for black hole (BH) growth. If sufficient 
matter becomes available very close to the nuclear BH and is accreted,
the nucleus of the galaxy will shine with enormous power as an AGN. 
In parallel, and probably before the fully
developed AGN phase, the merger will trigger a high SFR leading to a
luminous IR-active phase, either as 
a luminous IR galaxy LIRG or a ULIRG (see e.g., the simulations of
\cite{Hopkins2008}). \cite{Sanders1988} proposed such an evolutionary
scenario between ULIRGs and quasars almost  twenty five years ago. 
\cite{Kormendy2011} (and references
therein) proposed  an alternative to this merger-driven sequence,
where BHs in bulgeless galaxies and in
galaxies with pseudobulges probably grow through phases of low level
Seyfert-like activity. This growth is likely to be  driven 
stochastically by local processes (secular processes) not associated
with major mergers, and thus they would not have a global impact on the host
galaxy structure. 

Since the triggering of the activity in local ULIRGs and likely also in the
most luminous local LIRGs is driven by 
major mergers, ULIRGs tend to host more luminous AGN
(quasar or nearly quasar-like luminosities), 
slightly more massive BH and have higher AGN bolometric contributions
that local LIRGs (\cite{Veilleux2009, Nardini2010, 
AAH12,AAH13}). While the simplicity of the \cite{Sanders1988} scenario
is appealing, several works have argued that some modifications are needed. For
instance, based on the scatter of some of the properties of ULIRGs (e.g., AGN
contribution, Eddington ratios, obscuration) \cite{Veilleux2009}
proposed a revised evolutionary picture where the initial conditions
of the interaction would play an important role. Moreover,
\cite{Farrah2009} put forward an
evolutionary paradigm  of dusty AGN where 
some ULIRGs do not go through a quasar phase at all.

A number of works 
(see e.g., \cite{Sanders2004,AAH06})
have shown that major mergers are not prevalent in the population of
local LIRGs at $L_{\rm IR} \lesssim 3\times 10^{11}\,L_\odot$.
Therefore, in most local LIRGs the current episodes of star
formation activity and BH growth  could be due to other 
processes such as  minor 
mergers, fly-by companions, and/or secular evolution. Still, the same
sequence of events as in ULIRGs appears in LIRGs. 
\cite{AAH13} recently found that local LIRGs go through a 
distinct IR-bright star forming phase taking place prior to the bulk of
the current BH growth (that is, the AGN phase). In other words, there
is a time delay between the peaks of SF and BH growth, as
found for optically identified Seyfert galaxies in the local universe
(\cite{Davies2007,Wild2010}) and is postulated for local ULIRGs.  

The author wishes to thank the organizers of the workshop for the
invitation to give this review and also to M. Imanishi for sharing
his results for Figure~1. This work was supported in part by
the Augusto Gonz\'alez Linares Program
from the Universidad de Cantabria.

\end{document}